\documentclass[conference]{IEEEtran}
\usepackage{amsmath,amssymb,amsfonts}
\usepackage{algorithmic}
\usepackage{comment}
\usepackage{graphicx}
\usepackage{caption}
\usepackage{subcaption}
\usepackage{textcomp}
\usepackage{xcolor}
\usepackage{physics}
\usepackage[switch]{lineno}
\usepackage[normalem]{ulem}
\usepackage[section]{placeins}
\def\BibTeX{{\rm B\kern-.05em{\sc i\kern-.025em b}\kern-.08em
    T\kern-.1667em\lower.7ex\hbox{E}\kern-.125emX}}
\usepackage[separate-uncertainty,multi-part-units=single]{siunitx}
\usepackage[backend=biber,      
    style=nejm,
    articledoi=true,
        doi=true,                  
        url=false,                  
        isbn=false,                 
        natbib=true,
]{biblatex}
\title{\huge Muscle pre-stimulation tunes viscous-like perturbation rejection in legged hopping}

\author{Fabio Izzi$^{1,2}$, An Mo$^{2}$, Syn Schmitt$^{3}$, Alexander Badri-Spr\"owitz$^{2}$, Daniel F. B. Haeufle$^{1}$
\\
\IEEEauthorblockA{$^{1}$Hertie-Institute for Clinical Brain Research, University of T\"ubingen, T\"ubingen, Germany}
\IEEEauthorblockA{$^{2}$Dynamic Locomotion Group, Max Planck Institute for Intelligent Systems, Stuttgart, Germany}
\IEEEauthorblockA{$^{3}$Institute for Modelling and Simulation of Biomechanical Systems, University of Stuttgart, Stuttgart, Germany}
}

\begin{document}
\maketitle
%
\begin{abstract}
Muscle fibres possess unique visco-elastic properties, capable of generating a stabilising zero-delay response to unexpected perturbations.
This instantaneous response--- termed ``preflex''---is crucial in the presence of neuro-transmission delays, which are particularly hazardous during fast locomotion due to the short stance duration.
While the elastic contribution to preflexes has been studied extensively, research on the role of fibre viscosity due to the force-velocity relation remains unexplored.
Moreover, muscle models predict conditions with saturated force-velocity relations resulting in reduced viscous-like fibre engagement.
The goal of our study is to isolate and quantify the preflex force produced by the force-velocity relation. We use our approach to analyse two perturbed vertical hopping conditions, differing in their viscosity engagement at touch-down.
Both cases showed stable hopping patterns despite the reduced viscous-like response to ground perturbations in the saturated case. Moreover, the force-velocity relation was not the predominant factor driving energy adjustment to disturbance intensity.
From a robotics perspective, our results suggest that already a simple, constant damper mounted in parallel to an actuator could provide stabilising preflexes at and shortly after impact.\\ \\
\textbf{Keywords}: damping, muscle, legged locomotion, energy dissipation, passive adaptation, ground disturbance.
\end{abstract}

\section{Introduction}
\begin{figure*}[tb]
    \centering
    \includegraphics[trim={0cm 0cm 0cm 0cm},clip,width=\textwidth]{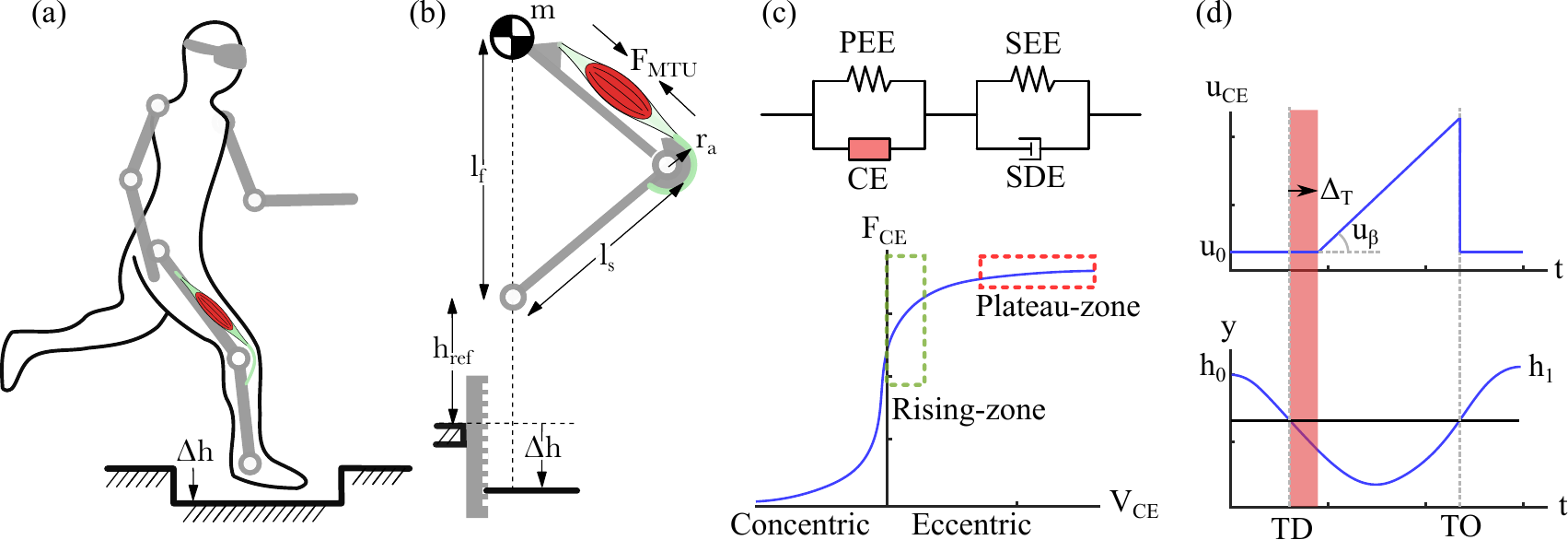}
    \caption{(a) Step perturbations are especially harmful during agile locomotion because neuro-transmission delay affects a considerable fraction of the stance duration. Because of their instantaneous response to ground disturbances, the mechanical properties of muscle-tendon units may be critical to prevent falling. (b) We simulate vertical hopping of a massless, two-segment leg model to investigate the stabilising response of the muscle fibre's force-velocity relation to step disturbance $\Delta h$. Here, $h_{ref}$ indicate the unperturbed periodic hopping height. (c) For our simulations, we use the Hill-type muscle model developed in \citep{Haeufle2014}, which includes a contractile element (CE), a parallel elastic element (PEE), a serial elastic element (SEE), and a serial damper element (SDE). The force-velocity relation models the dependence of the muscle fibre force $F_{CE}$ on the muscle fibre velocity $v_{CE}$. We divided the eccentric side ($v_{CE} > 0$ \SI{}{\m/s}) into two zones to explore the stabilising influence of the force-velocity state at touch-down. Inside the plateau-zone, the force-velocity relation is mostly flat, and hence little variation in muscle fibre force $F_{CE}$ occurs as a result of variation in muscle fibre velocity $v_{CE}$. The opposite is true within the rising-zone. (d) To stimulate the muscle and generate periodic hopping, we apply a delayed ramp signal, with prestimulation value $u_0$ and slope $u_{\beta}$. During the preflex period $\Delta_T=\SI{30}{ms}$ post-ground impact (TD) the stimulation is constant. This preflex period is the focus of our investigation. $h_0$ indicates an initial drop height and $h_1$ the corresponding return apex. At take-off (TO), the neuronal stimulation resets to its initial prestimulation value $u_0$.}
    \label{fig:model}
\end{figure*}

Muscles contribute to the control of biological movements with visco-elastic properties \citep{John_2013,Krogt_2009,Haeufle_2020a,Haeufle_2020b,Patla_2003}.
These properties allow muscles to react instantly to unexpected perturbations, producing a zero-delay response known as ``preflex'' \citep{Loeb1995}.
Generating preflexes is particularly advantageous during rapid movements, when neuro-transmission delays limit strategies from the central nervous system and peripheral reflexes \citep{Patla_2003,More2018}.
A specific muscle property which is relevant for the generation of preflexes is their asymmetric capacity to produce and dissipate power between shortening and extending muscle fibres, i.e., concentric and eccentric contraction, respectively \citep{Siebert2014,Biewener1998,Herzog2018}. Interestingly, muscles generate less positive than negative power \citep{Herzog2018}, suggesting they may possess inherent damping properties which may be exploited against disturbances.
In macroscopic muscle models, this power production asymmetry is captured by a phenomenological hyperbolic force-velocity relation attributed to the muscle fibres (Fig.\ref{fig:model}) \citep{Soest1993a, Haeufle2014}.

Simulations of musculoskeletal models have shown that the force-velocity relation contributes to the robustness of hopping and walking \citep{Haeufle2010,Gerritsen1998}. By embedding non-linear viscosity in a muscle-driven system \citep{Geyer2003,Haeufle2010}, the force-velocity relation is expected to mitigate sudden perturbations without additional control load on the central nervous system. As such, the force-velocity relation could play a major role in the muscle preflex response during fast, agile locomotion. For example, it may facilitate the quick rejection of perturbations by instantly adapting the muscle force to changes in impact velocity due to a different ground height, similarly to what has been observed when implementing a parallel viscous damper in a legged system \citep{Mo2020,Heim2020,Abraham2015}.

According to \citep{Joyce1969,Till2008}, however, the force-velocity relation may plateau at high fibre-stretching velocities, implying a potentially undesired decrease of stabilising capacity in severely perturbed conditions. It is still unclear how and to what degree the force-velocity relation can contribute to the preflex response during perturbed agile locomotion.

In this study, we quantify the contribution of the muscle fibres' force-velocity relation to preflex during perturbed vertical hopping. The novelty of our analysis is dual: on the one hand, we present an explicit quantification of the force component and dissipated energy generated by the force-velocity relation and compare them to the contributions from muscle elasticity and muscle activity; on the other hand, we examine how the stabilising response of the force-velocity relation depends on its operational state at touch-down caused by different levels of neuronal stimulation prior to impact. Understanding the inherent stabilising properties of neuro-musculoskeletal mechanics is of important guidance to discover new design principles for more robust and easier-to-control bio-robotic applications.

\section{Methods}
\subsection{Musculoskeletal model}
For our study, we used a modified version of the musculoskeletal model developed in \citep{Geyer2003}, with identical kinematics but a revised muscle-tendon unit \citep{Haeufle2014}. The model consisted of a two-segment leg with total mass lumped at the hip and motion constrained to the vertical axis (Fig.\ref{fig:model}b).
The knee joint was actuated by an extensor muscle-tendon unit, whose force outcome $F_{MTU}$ produced a torque at the knee joint equal to:
\begin{equation}
T_{knee} =
  \begin{cases}
    r_{a} F_{MTU}  & \quad \text{, in stance}\\
    0 & \quad \text{, in flight}
  \end{cases}
\end{equation}
with $r_{a}$ being the muscle lever arm. During the flight phase the leg geometry was fixed, with leg length equal to $l_f$. For this reason, we excluded those hopping conditions with return apex $h_1 < l_f$, for which a reset of the leg geometry would have been impossible during continuous hopping. Notice that, however, we limited each simulation to one hopping cycle. Stance started when the mass vertical position $y \leq l_f$, and the following take-off occurred when either $y>l_f$ or the ground reaction force $F_{leg} \leq \SI{0}{N}$, given that we defined positive $F_{leg}$ to push the mass upwards.

For our muscle-tendon unit we used the Hill-type model developed in \citep{Haeufle2014}. To summarise, it consists of four elements, as shown in Fig.\ref{fig:model}c: a contractile element (CE), a parallel elastic element (PEE), a serial elastic element (SEE), and a serial damping element (SDE). These four components fulfil the force equilibrium
\begin{equation}\label{eq:FmtuBalance}
    F_{MTU} = F_{CE} +  F_{PEE} = F_{SEE}  + F_{SDE}
\end{equation}
The contractile element represents the collective contribution of the muscle fibres to the muscle contraction. In our model, the parallel elastic element (PEE) engages for fibre lengths $l_{CE}>95\%$ of the muscle fibres' optimal length $l_{opt}$. As such, $F_{CE}$ represents the net force generated by the muscle-tendon unit $F_{MTU}$ for the majority of our simulated conditions.

The contractile element force $F_{CE}$ is a non-linear function of the fibre velocity $v_{CE}$ (force-velocity relation), fibre length $l_{CE}$ (force-length relation), and muscle activity $a$ which is in turn dependent on the neuronal stimulation $u$ received by the muscle fibres.
The force-velocity relation comprises two regions, as shown in Fig.\ref{fig:model}c: the concentric contraction and eccentric contraction regions, respectively characterised by shortening ($v_{CE} \leq \SI{0}{\m/s}$) and stretching ($v_{CE} > \SI{0}{\m/s}$) of muscle fibres. The initial response to ground impact occurs with muscle fibres in eccentric contraction, which made the eccentric region of the force-velocity relation the main focus of our study.

Within this region, we separated two zones of interest: the rising-zone and the plateau-zone. In rising-zone, a deviation in fibre velocity produces a large deviation in $F_{CE}$, while $F_{CE}$ mostly saturates along the plateau-zone. Therefore, we expected the muscle fibres to adapt the muscle response to ground perturbations mainly when operating within the rising-zone at touch-down.

\begin{figure*}[htp]
    \centering
    \includegraphics[trim={0cm 0cm 0cm 0cm},clip,scale=1]{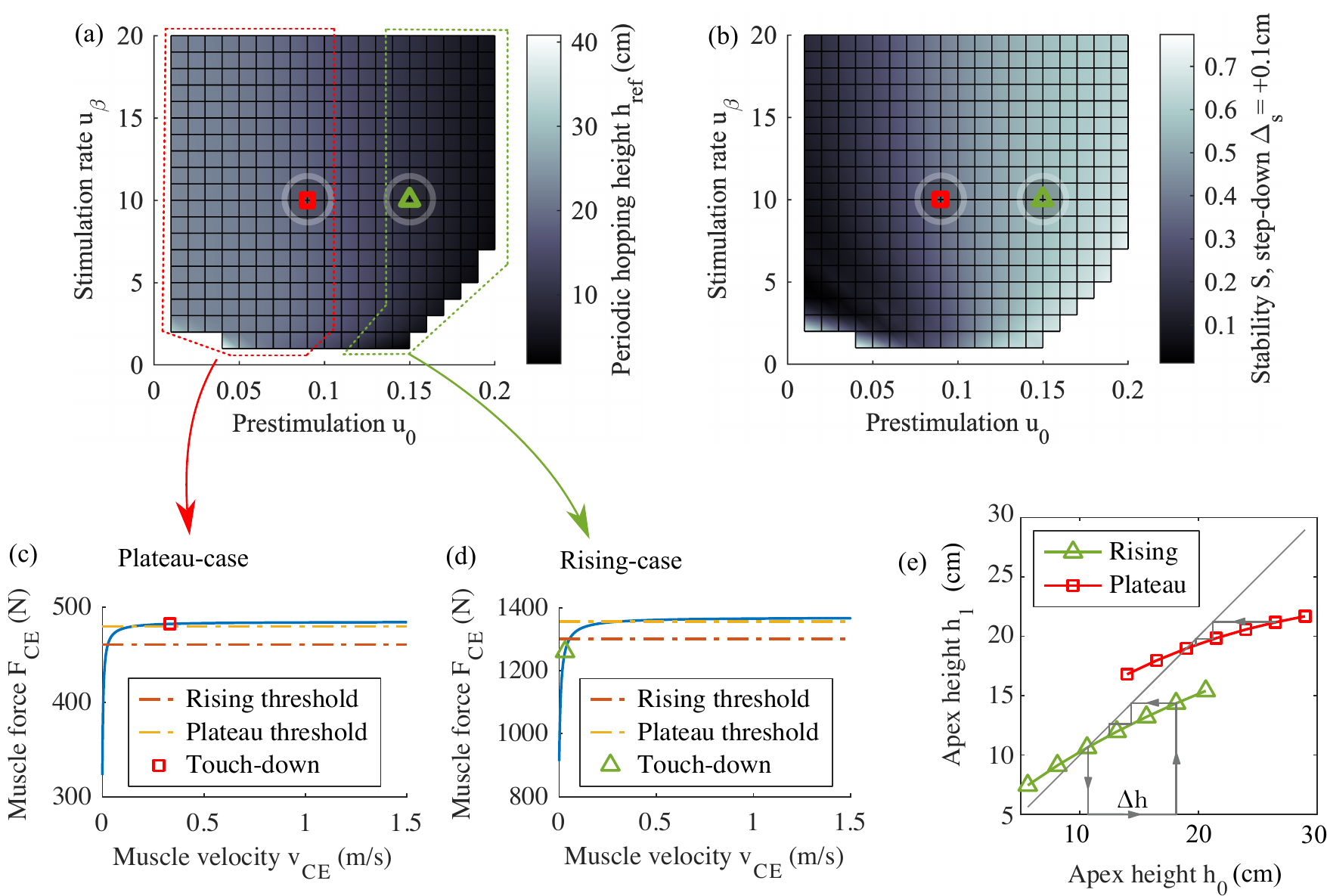}
    \caption{(a) Periodic hopping heights $h_{ref}$ for the stimulation input pairs ($u_0,u_{\beta}$). Enclosing lines separate the plateau-cases from the rising-cases, i.e., conditions at touch-down in the plateau-zone and rising-zone, respectively. The square and triangle symbols denote the plateau- and rising-cases, respectively, which we selected for further investigation. (b) Stability analysis for step-down disturbance ($\Delta_s = $\SI{+0.1}{\cm}). Stability analysis for step-up disturbance ($\Delta_s = $ \SI{-0.1}{\cm}) yielded similar results. (c-d) Shape of the force-velocity relation at touch-down for the selected plateau-case and rising-case. Dashed lines indicate the two thresholds used to differentiate the plateau- and rising-zone. The markers (square and triangle) indicate the specific force-velocity state at touch-down. (e) Apex return map for the selected plateau- and rising-case, with $h_0$ indicating each starting apex height, and $h_1$ the associated return apex after one hopping cycle. A more horizontal inclination angle around the intersection point with the diagonal denotes a faster rejection of a specific ground perturbation $\Delta_h$.}
    \label{fig:periodicHopping}
\end{figure*}

We defined the muscle fibres to operate in the rising- or plateau-zone at touch-down as in:
\begin{equation}
  \begin{cases}
    \text{rising-zone}  & \quad \text{if } F_{CE}(t_{TD}) < 0.95\,\tilde F_{CE}\\
    \text{plateau-zone}  & \quad \text{if } F_{CE}(t_{TD}) > 0.99\,\tilde F_{CE}
  \end{cases}
  \label{eq:RPdefinition}
\end{equation}
where $t_{TD}$ is the touch-down time, and \(\tilde F_{CE}\) is the force that the contractile element produces if $v_{CE} = \SI{1.5}{\m/s}$ at ground impact. As such, \(\tilde F_{CE}\) is the largest expected force produced by the contractile element for fibre length values and neuronal stimulation measured at touch-down. Fig.\ref{fig:periodicHopping} further visualises our classification approach.

\subsection{Muscle stimulation}
We stimulated the muscle fibres using a delayed ramp signal as neuronal stimulus $u(t)$ (Fig.\ref{fig:model}d). This stimulation was triggered by the touch-down event and transmitted in an open-loop fashion, as described in:
\begin{equation}
\footnotesize
    u(t) = \begin{cases}
      u_0 + u_\beta \, \left(t - t_{TD} - \Delta_T \right) & \quad \text{, for } t_{TD} + \Delta_T < t < t_{TO} \\
      u_0  & \quad \text{, otherwise}
    \end{cases}
\end{equation}
where $u_0$ is the muscle prestimulation level, $u_\beta$ the slope of the ramp stimulus, $t_{TD}$ and $t_{TO}$ the touch-down and take-off time-events, respectively, and $\Delta_T$ the neuro-transmission delay.

Before reaching muscle fibres, the neuronal signal $u(t)$ is converted into muscle activity $a(t)$ by the activation dynamics, which consists of a first order differential equation that takes into account the fibre length dependency, as described in \citep{rockenfeller2015,hatze1977}. The activation dynamics limits the muscle activity between $[0,1]$.

\subsection{Muscle fibre force decomposition}
The aim of our investigation was to quantify the contribution of the force-velocity relation to the force produced by the muscle fibres around impact, and separate it from the contribution of fibre elasticity and muscle activity.
In our hopping scenario, the muscle fibres exert a force $F_{CE}$ on the leg geometry:
\begin{equation}
  F_{CE}(t) =
  \begin{cases}
    F_{CE}^{0}  & \text{, in flight}\\
    F_{CE}^{0} + f_{CE}\left(v_{CE}\left(t\right),l_{CE}\left(t\right),a\left(t\right)\right) & \text{, in stance}
  \end{cases}
  \label{eq:Fce}
\end{equation}
where $F_{CE}^0$ is the force pretension due to the muscle fibres' prestimulation. In the following, we will omit the time-dependence for better readability. Eq.\ref{eq:Fce} shows that muscle fibres produce a force during stance that is a non-linear function of the muscle fibre velocity $v_{CE}$, length $l_{CE}$, and muscle activity $a$. It is possible to separate these three contributions by interpreting $f_{CE}$ as follows:
\begin{eqnarray}
    f_{CE} &=& \int{\frac{df\left(v_{CE},l_{CE},a\right)}{dt} \, dt}\nonumber\\
    &=& \int{\left(\frac{\partial f_{CE}}{\partial v_{CE}}\frac{dv_{CE}}{dt} + \frac{\partial f_{CE}}{\partial l_{CE}}\frac{dl_{CE}}{dt} + \frac{\partial f_{CE}}{\partial a}\frac{da}{dt} \right) \,dt} \nonumber\\
    &=& \int{\frac{\partial f_{CE}}{\partial v_{CE}}dv_{CE}} + \int{\frac{\partial f_{CE}}{\partial l_{CE}}dl_{CE}} + \int{\frac{\partial f_{CE}}{\partial a} da}
    \label{eq:fcmpcont}
\end{eqnarray}
From eq.\ref{eq:fcmpcont} and eq.\ref{eq:Fce}, we can separate the muscle fibre force during stance into four components:
\begin{equation}
  \begin{aligned}
    F_{CE} &= F^0_{CE} + F^V_{CE} + F^L_{CE} + F^A_{CE} \\
    F^V_{CE}&= \int{\frac{\partial f_{CE}}{\partial v_{CE}}dv_{CE}}\\
    F^L_{CE} &= \int{\frac{\partial f_{CE}}{\partial l_{CE}}dl_{CE}}\\
    F^A_{CE} &= \int{\frac{\partial f_{CE}}{\partial a} da}
\end{aligned}
        \label{eq:fcmp}
\end{equation}
where $F_{CE}^V$ is the force contribution produced by the force-velocity relation, $F_{CE}^L$ from the force-length relation, and $F_{CE}^A$ from the muscle activity, in turn associated with the stimulation signal received by muscle fibres.

Given a mathematical formulation of the contractile element dynamics $f_{CE}$, partial derivatives in eq.\ref{eq:fcmp} can be solved analytically: for our analysis we implemented $f_{CE}$ as described in \citep{Haeufle2014}. We numerically solved each integral term in eq.\ref{eq:fcmp} during data post-processing (trapezoidal rule--- \emph{cumtrapz}, Matlab R2018a, Mathworks Inc., Natick, MA, USA), using the stance duration as integration interval and the same variable step size as in the associated simulation. Since touch-down ($t=t_{TD}$) occurs instantaneously in our model, the initial values for the numerical integration of eq.\ref{eq:fcmp} can be algebraically computed:
\begin{gather}
  F_{CE}^L(t_{TD}) = F_{CE}^A(t_{TD}) = 0 \nonumber\\
  F_{CE}^V(t_{TD}) =  F_{CE}(t_{TD}) - F_{CE}^0
  \label{eq:FceDecTD}
\end{gather}
Eq.\ref{eq:FceDecTD} shows that the variation in muscle fibre force observed at touch-down can only come from the contribution of force-velocity relation, since muscle fibre geometry ($l_{CE}$) and activity ($a$) remain constant across an instantaneous ground impact.

\subsection{Experimental procedure}
First, we identified periodic hopping conditions for a grid of model inputs ($u_0 = [0.01:0.01:0.2]$, $u_\beta = [1:1:20]$). For each ($u_0$, $u_\beta$) combination, we computed the drop height ($h_{ref}$) producing the smallest variation in following apex height ($h_1$) through optimisation--- $min(h_{1} -h_{ref})^2$ (\emph{fmincon}, Matlab R2018a). In our analysis, we only included input conditions producing periodic hopping, which we defined as $h_{ref} > \SI{1}{\cm}$ and $\abs{h_{1}-h_{ref}} < \SI{0.01}{\mm}$.

We measured the stability ($S$) of each periodic hopping condition by simulating one perturbed hopping cycle ($h_{ref} + \Delta_s$) and then computing
\begin{equation}
S = \frac{h_1^* - h_{ref}}{\Delta_s}
\label{eq:stability}
\end{equation}
where $h_1^*$ is the perturbed return apex height, and $|S|<1$ guarantees stable periodic hopping. For completeness, we tested bidirectional perturbations per each periodic hopping condition ($\Delta_{s} = \pm \SI{0.1}{cm}$). We further classified the periodic hopping conditions based on the zone of force-velocity relation in which they operate at touch-down, as defined in eq.\ref{eq:RPdefinition}.

Finally, we selected two settings, one rising-case and one plateau-case, and investigated the muscle fibres' mechanical response to ground perturbations, during the \SI{30}{\ms}-lasting preflex time window, immediately after touch-down. We perturbed these two selected cases with step perturbations $\Delta_h = [-5,-2.5,0,2.5,5,7.5,10]$ cm, for a total of two step-up perturbations (negative sign), and four step-down perturbations (positive sign). Considering that $l_f = \SI{0.99}{\meter}$, this corresponds to ground perturbations within the range of about [-5,10]\% of the reference leg length $l_f$.

\subsection{Simulation platform and parameters}
We implemented our model in Simulink R2018a (Mathworks Inc., Natick, MA, USA). As numerical solver we used ode45 with maximum step size of $10^{-4}$ (absolute and relative error tolerances of $10^{-8}$). The main model parameters are listed in TABLE \ref{tab:modelPar}.

\begin{table}[]
    \centering
\caption{Model parameters, adapted from \citep{Geyer2003}.}
\begin{tabular}{ll}
\hline & \\
parameter & value \\
\hline & \\
body weight $m$ & $80 \mathrm{~kg}$ \\
gravitational constant $g$ & $9.81 \mathrm{~m}/\mathrm{s}^{2}$ \\
assumed flight leg length $\ell_{\mathrm{f}}$ & $0.99 \mathrm{~m}$ \\
segment length $\ell_{\mathrm{s}}$ & $0.5 \mathrm{~m}$ \\
optimum length $\ell_{\text {opt }}$ & $0.1 \mathrm{~m}$ \\
lever arm $r_{a}$ & $0.04 \mathrm{~m}$ \\
maximum isometric force $F_{\max }$ & $22 \mathrm{~kN}$ \\
neuro-transmission delay $\Delta_{T}$ & $0.03 \mathrm{~s}$ \\
& \\
\hline
\end{tabular}
    \label{tab:modelPar}
\end{table}

\section{Results}
\subsection{Periodic hopping analysis}
Out of 400 tested conditions, we found 381 $\left(u_0,u_\beta\right)$ input pairs producing periodic hopping cycles (Fig.\ref{fig:periodicHopping}a), which were all stable (Fig.\ref{fig:periodicHopping}b). Stable periodic hopping was possible with the touch-down state $\left(F_{CE},v_{CE}\right)$ anywhere on the eccentric side of the force-velocity relation: either within the plateau-zone, rising-zone, or in-between (Fig.\ref{fig:periodicHopping}a).
Fig.\ref{fig:periodicHopping}a shows a sharp clustering of the periodic hopping conditions according to the force-velocity state at touch-down (plateau-case vs. rising-case enclosing lines). This distribution almost exclusively depends on the muscle prestimulation level $u_0$. Plateau-cases were characterised by larger periodic hopping heights than the rising-cases (\SI{21.2 \pm 2.5}{\cm} and \SI{7.9 \pm 2.6}{\cm}, mean $\pm$ st.d. respectively).
In contrast to our expectation, conditions belonging to the plateau-zone were also more stable  than those belonging to the rising-zone, as the associated $S$ values were closer to zero (mean $\pm$ st.d.: $0.25\pm0.13$ for plateau-cases vs. $0.60\pm0.04$ for rising-cases).

\subsection{Preflex response analysis}
To investigate the preflex response of the muscle fibres to ground perturbations, we selected two periodic hopping conditions, one with touch-down fibre velocity inside the plateau-zone of the force-velocity relation ($u_0 = 0.09$, $u_{\beta} = 10$), and one with touch-down fibre velocity inside the rising-zone ($u_0 = 0.15$, $u_{\beta} = 10$) (Fig.\ref{fig:periodicHopping}c,d). For clarity, we will refer to these two experimental conditions as plateau-case and rising-case, respectively, and, as they represent unperturbed periodic hopping, we further classify them as reference cases.

The plateau-case features a 18.97 cm apex height $h_{ref}$ and a 338 ms stance duration, while the rising-case has a 10.61 cm apex height $h_{ref}$ and a 255 ms stance duration. Muscle pretension force $F_{CE}^{0}$ values were \SI{323}{\newton} and \SI{913}{\newton} for the plateau- and rising-case, respectively. Each reference case was tested against ground perturbations $\Delta_h = [-5,-2.5, 0, 2.5, 5, 7.5, 10]$ cm, for a total of two step-up (negative sign) and four step-down perturbations (positive sign). Inspection of the apex return map confirms stronger asymptotic stability for the plateau-case (Fig. \ref{fig:periodicHopping}e). Leg compression during the stance duration was never sufficient to engage the parallel elastic element PEE in either the plateau- or rising-case, perturbed or not. As such, the muscle fibre force $F_{CE}$ was representative of the net muscle-tendon-unit response for our selected conditions (see eq. \ref{eq:FmtuBalance}).

\begin{figure*}[htb]
    \centering
    \includegraphics[trim={0cm 0cm 0cm 0cm},clip,scale=1]{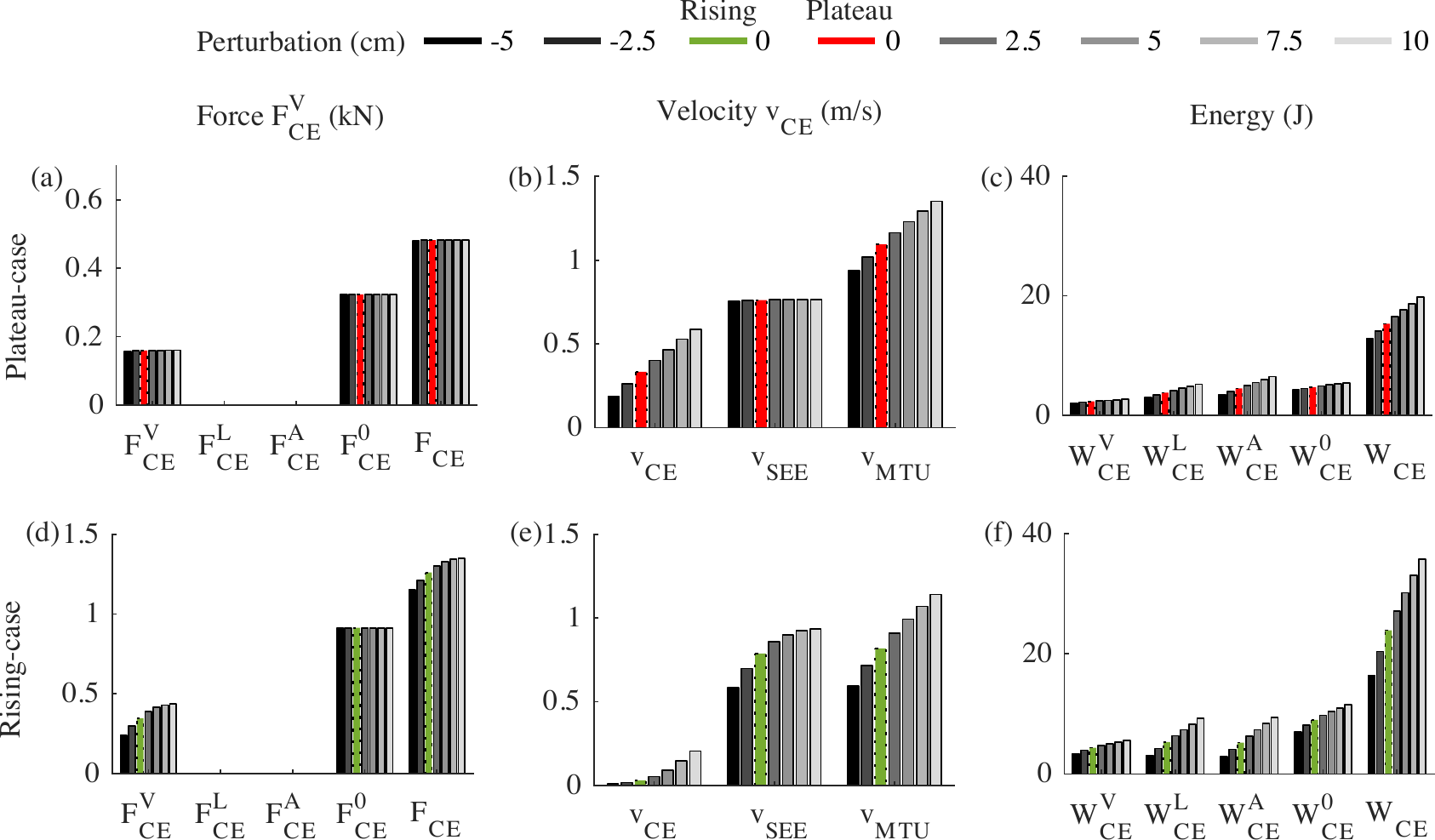}
    \caption{(a,d) Muscle fibre force at touch-down, separated into the contribution of: the force-velocity relation ($F_{CE}^V$), force-length relation ($F_{CE}^L$), muscle activity ($F_{CE}^A$), and muscle prestimulation ($F_{CE}^0$). For the sake of comparison, $F_{CE}$ is the net force produced by the muscle fibres at touch-down ($F_{CE} = F_{CE}^V + F_{CE}^L + F_{CE}^A + F_{CE}^0$). (b,e) Touch-down muscle fibre velocity ($v_{CE}$), tendon velocity ($v_{SEE}$) and muscle-tendon unit velocity ($v_{MTU} = v_{CE} + v_{SEE}$). (c,f) Energy dissipated by the muscle fibres throughout the preflex duration, separated into the contribution of: the force-velocity relation ($W_{CE}^V$), force-length relation ($W_{CE}^L$), muscle activity ($W_{CE}^A$), and muscle prestimulation ($W_{CE}^0$). For the sake of comparison, $W_{CE}$ is the net energy dissipated by muscle fibres ($W_{CE} = W_{CE}^V + W_{CE}^L + W_{CE}^A + W_{CE}^0$). The results illustrate the plateau-case (a-c) and rising-case (d-f) under different levels of ground perturbation.}
    \label{fig:bars}
\end{figure*}

Across all the tested conditions, the force-velocity relation produced an instantaneous, breaking force at touch-down ($F_{CE}^{V}$ in Fig. \ref{fig:bars}a,d).
This was expected because a massless leg and instantaneous ground impact generate a jump in muscle-tendon velocity $v_{MTU}$, and thus muscle fibre velocity $v_{CE}$, while muscle geometry and stimulation level stay constant across touch-down (see eq.\ref{eq:FceDecTD}). The jump in muscle force at touch-down was substantial: during unperturbed hopping, $F_{CE}^V$ was equivalent to approximately $49\%$ (plateau-case) and $38\%$ (rising-case) of the original values of muscle pretension $F_{CE}^0$.

To check for a possible viscous-like response due to the force-velocity relation, we looked at the variation of $F_{CE}^{V}$ across the perturbed impact conditions. For the rising-case, the touch-down value of $F_{CE}^{V}$ adapted to the ground perturbation intensity, ranging from \SI{238}{\newton} to \SI{436}{\newton} ($26\%$ and $48\%$ of the original muscle pretension). However, a similar adaptation was absent for the plateau-case. This finding confirms our hypothesis that the initial adaptation to ground perturbation produced by the muscle fibres depends on the operative zone of the force-velocity relation at touch-down.

However, the adaptation of $F_{CE}^{V}$ observed for the rising-case was not proportional to the variation of muscle fibre velocity $v_{CE}$ across ground perturbations, as one would expect from an ideal viscous response (Fig. \ref{fig:bars}d vs. e). While the $v_{CE}$ increments grew in size with higher drop heights, the corresponding $F_{CE}$ increments lessened. These two opposite trends imply that incremental impact velocities move the touch-down state of the force-velocity relation towards the plateau-zone, thereby reducing the touch-down adaptation of $F_{CE}^V$.

Impact transmission to the muscle fibres differed between the plateau- and rising-case. During the plateau-case, $v_{CE}$ dependence on the step perturbation intensity closely matched the trend of $v_{MTU}$, while $v_{SEE}$ remained about constant across conditions (Fig.\ref{fig:bars}b). This means that in the plateau-case, the change in impact condition almost exclusively affects the muscle fibre velocity and not the tendon velocity. In comparison, during the rising-case, $v_{MTU}$ showed a profile more similar to that of $v_{SEE}$, especially during step-up and mild step-down perturbations (Fig.\ref{fig:bars}e). This means that the tendon stretching velocity was now the most affected by the ground perturbation at touch-down. Overall, these results demonstrate that impact velocity and muscle fibre velocity at touch-down are not directly proportional. Rather, there is a complex mapping that depends on the operating conditions of the force-velocity relation and on the level of prestimulation. This further supports our hypothesis that the different operating-zones of the force-velocity relation affect the muscle fibres' behaviour.

In spite of the differences observed at touch-down, both the plateau-case and rising-case produce stabilising responses within the preflex duration, as indicated by the variation in total dissipated energy $W_{CE}$ as a function of the ground perturbation intensity (Fig.\ref{fig:bars}c,f). The highest step-up and step-down perturbations, when applied to the periodic hopping condition, resulted in a change in total dissipated energy $W_{CE}$ of \SI{-2.50}{\joule} and \SI{4.43}{\joule} for the plateau-case, and \SI{-7.49}{\joule} and \SI{11.85}{\joule} for the rising-case, respectively. This means that by the end of the preflex duration, the plateau-case rejected around 6.4\% and 5.6\% of the change in system energy induced by the biggest step-up and step-down perturbations, respectively, whereas the rising-case rejected 19.1\% and 15.1\%. For reference, the preflex duration (\SI{30}{\ms}) during unperturbed hopping represented 8.9\% and 11.8\% of the stance phase in the plateau-case and rising-case, respectively. This indicates that in terms of net dissipated energy during preflex ($W_{CE}$), muscle fibres rejected ground perturbations better in the rising-case.

The force-velocity relation, along with the force-length relation, muscle pretension, and even activation dynamics, all contributed to counter the step perturbation during the preflex phase in the form of negative work. It is worth noting that the energetic contribution from the muscle activity stimulation $W_{CE}^A$ is not zero during preflex, despite the stimulation signal being constant ($u=u_0$). Our activation dynamic model includes an internal muscle-length dependency, which alters muscle activity and force due to fibres length changes, even with a constant stimulation signal. Contrary to what we expected, the energy dissipated due to the force-velocity relation $W_{CE}^V$ contributed the least to the total dissipated energy $W_{CE}$ across almost all examined scenarios. Also, the change of $W_{CE}^V$ as a function of the ground perturbation intensity contributed the least to the observed adaptation of $W_{CE}$ to the perturbation level, as most of the change in $W_{CE}$ came from the force-length relation ($W_{CE}^L$) and activity induced changes ($W_{CE}^A$) (Fig.\ref{fig:bars}c,f). Yet, the rising-case showed a more than threefold adaptation of $W_{CE}^V$ to the ground perturbation intensity in relation to the plateau-case, ranging from \SI{3.4}{\joule} to \SI{5.6}{\joule} compared to \SI{2.1}{\joule} and \SI{2.7}{\joule}, respectively.

\begin{figure*}[htb]
    \centering
    \includegraphics[trim={0cm 0cm 0cm 0cm},clip,scale=1]{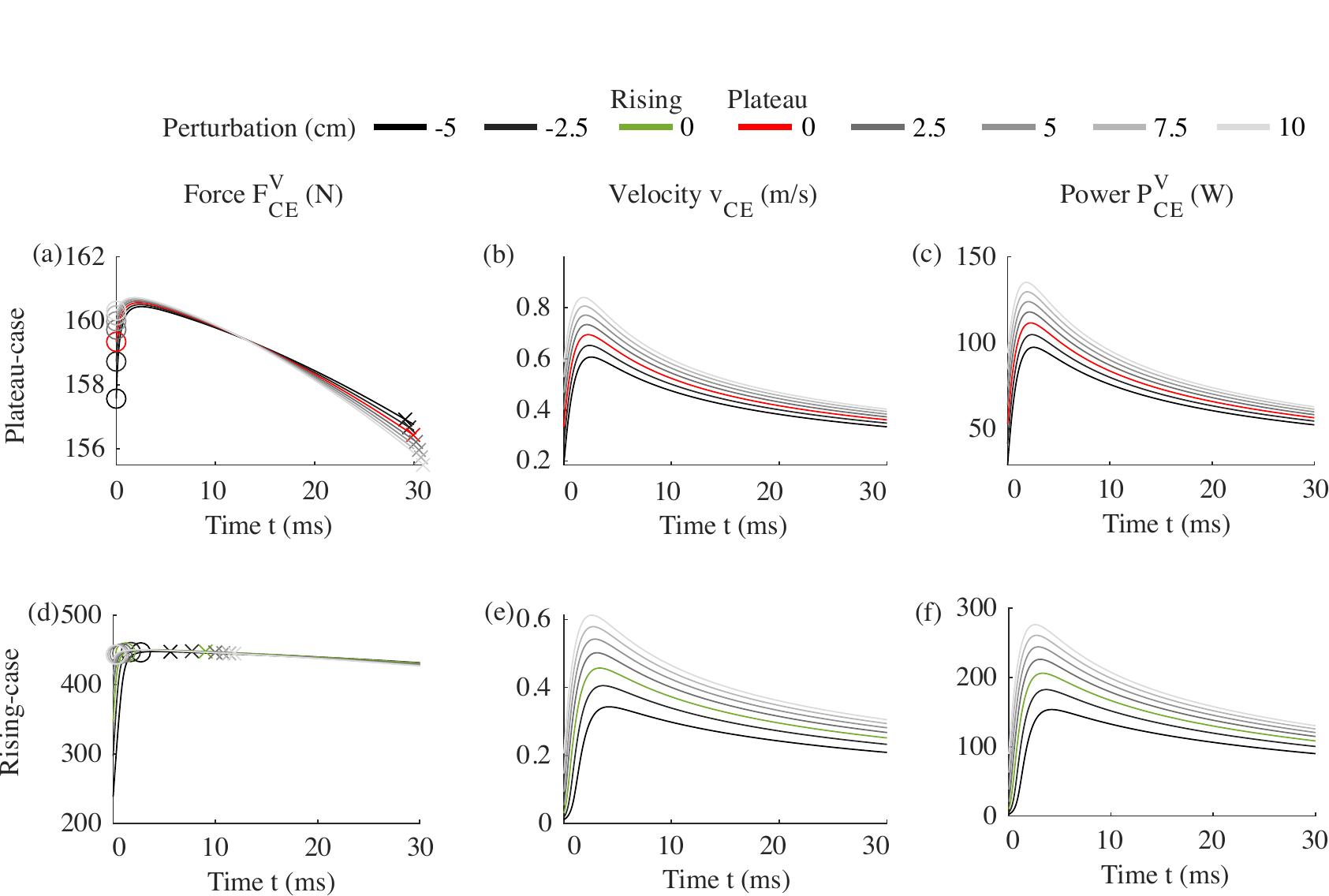}
    \caption{(a,d) Time progression of the force contribution from the force-velocity relation $F_{CE}^V$. Circles indicate the entry of the muscle state ($F_{CE},v_{CE}$) inside the plateau-zone; crosses indicate the exit from the plateau-zone. In no case did ($F_{CE},v_{CE}$) return to the rising-zone for the remaining preflex duration. (b,e) Time progression of the muscle fibre velocity $v_{CE}$. (c,f) Time progression of the power $P_{CE}^V = F_{CE}^V \cdot v_{CE}$. Results illustrate the plateau-case (a-c) and the rising-case (d-f) from touch-down to the end of the preflex duration, and in response to various ground perturbation levels.}
    \label{fig:time_variables}
\end{figure*}

To further understand how the force-velocity relation dissipates energy during preflex, we inspected the time progression of the force $F_{CE}^V$, fibre velocity $v_{CE}$ and associated power $P_{CE}^V$ (Fig.\ref{fig:time_variables}). For both the plateau- and the rising-case, $F_{CE}^V$ trajectories show minimal dependence on the ground perturbation level (Fig.\ref{fig:time_variables}a,d). This occurs because the muscle fibre velocity $v_{CE}$ rapidly rises following an impact (Fig.\ref{fig:time_variables}b,e), causing the ($F_{CE}$, $v_{CE}$) state of the muscle to enter the plateau-zone of the force-velocity relation early after touch-down and remain outside the rising-zone throughout the majority of the preflex duration (markers in Fig.\ref{fig:time_variables}a,d). As a consequence, even in the rising-case, $F_{CE}^V$ trajectories exhibit little variability early after impact.

On the contrary, the trajectory of the fibre velocity $v_{CE}$ changes noticeably across ground perturbations (Fig.\ref{fig:time_variables}b,e), with trends similar to those of the power $P_{CE}^V$ (Fig.\ref{fig:time_variables}c,f). This means that changes in the velocity profile $v_{CE}$, rather than changes in the force profile $F_{CE}^V$, are primarily responsible for any adaptation of the energy $W_{CE}^V$ to the ground perturbation level. This observation is supported further by the observed adaptation of the energy dissipated by the muscle pretension ($W_{CE}^0$ in Fig.\ref{fig:bars}c,f), which only an adaptation of the $v_{CE}$ trajectory can explain.

\section{Discussion}
In this study, we quantified the regulatory effect of the Hill-type muscle fibres' force-velocity relation in response to ground perturbations during vertical hopping. We focused our analysis on the early stance response (preflex period $\Delta_T=\SI{30}{ms}$ after touch-down), during which a viscous-like response from the muscle fibres is expected to play a significant role in motion stabilisation \citep{Geyer2003,Haeufle2010}. We hypothesised (HP:1) that the contribution of the force-velocity relation to the perturbation rejection would depend on its initial state at touch-down. For this reason, we compared two different periodic hopping cases: during the plateau-case, the muscle fibre's force-velocity state at touch-down was within the saturated zone of the force-velocity relation, whereas during the rising-case, it was on the steep incline (Fig.\ref{fig:periodicHopping}c,d). Hence, we predicted the rising-case to produce a larger rejection of ground disturbances than the saturated plateau-case. Furthermore, we expected (HP:2) the force-velocity relation to be the major factor contributing to the leg's adaptation to the ground perturbation's intensity.

\subsection{HP:1 - Plateau-case and rising-case comparison}
During preflex, the rising-case exhibited a greater adaptation of the energy dissipated by the force-velocity relation ($W_{CE}^V$) to ground perturbation intensity than the plateau-case (Fig.\ref{fig:bars}, c vs. f), thus indicating a confirmation of hypothesis 1. However, this greater adaptation was not crucial for the hopping stability: we actually found the plateau-case to produce more stable hopping patterns than the rising-case (Fig.\ref{fig:periodicHopping}e).

A closer look at the origin of the viscous preflex adaptation $W_{CE}^V$ shows that the force-velocity component $F_{CE}^V$ converged shortly after touch-down to the same force-levels in both the plateau- and, more surprisingly, the rising-case (Fig.\ref{fig:time_variables}a,d).
This convergence is due to a rapid shift of the operative point of the force-velocity relation to the plateau-zone during the preflex: shortly after touch-down, a rapid increase in the stretching velocity of the muscle fibres causes the state of the force-velocity relation to leave the rising-zone and operate outside it for the majority of the preflex phase (mean $\pm$ st.d.: \SI{99.2 \pm 1.1}{\percent} and \SI{100(0)}{\percent} of the preflex duration for the rising- and plateau-case, respectively).
As a result, changes in the velocity trajectory due to different intensity of the step perturbation have soon little effect on the $F_{CE}^V$ profile. The fact that an increasing muscle fibre velocity pushes the $(F_{CE},v_{CE})$ state outside the rising-zone also explains why the rising-case adaptation of $F_{CE}^V$ at touch-down reduced with larger step-down perturbations (Fig.\ref{fig:bars}d vs e).
For the largest step-down perturbations ($\Delta_h = [5, 7.5, 10]$ cm) velocity at impact was already in the transition zone towards plateau-zone.
Therefore, our findings show that adaptation of $W_{CE}^V$ (Fig.\ref{fig:bars}c,f) was only partially due to a variation in the force component $F_{CE}^V$.

Our data rather suggest that the energy dissipated due to the force-velocity relation ($W_{CE}^V$) is mostly driven by the change of the velocity $v_{CE}$ trajectory (Fig.\ref{fig:time_variables}b,e). This implies an intrinsic regulation of the system energy due to natural adaptation of the system kinematics to the impact condition. As power is the product of velocity and force, controlling the magnitude of the force $F_{CE}^V$ permits the tuning of this intrinsic, velocity-driven, energy regulation. In consequence, a simple friction damper of controllable magnitude may be sufficient to regulate energy in a hopper robot, as long as the velocity profile can be adapted to variable ground impact conditions. This concept offers an intriguing new paradigm for controlling tuneable, physical damping in legged robots, providing an alternative solution to strategies exploiting more complex viscous-like characteristics, as simulated in \citep{Abraham2015, Heim2020} and implemented in \citep{Mo2020}.

It is worth noting that during preflex, the muscle fibres elongate and the activation state of the muscle-tendon unit increases due to the dependence of the activation dynamics on the muscle fibre length. In our muscle model, both these two factors contribute to amplify the curvature of the force-velocity relation, similarly to what is described in Fig.7G-H in \citep{Krogt_2009}. When the curvature of the force-velocity relation becomes steeper, the overall sensitiveness of the muscle fibre force $F_{CE}$ to the muscle fibre velocity $v_{CE}$ increases. Our results, however, suggest that muscle fibres' elongation during preflex, and the consequential increase in muscle activity, was insufficient to alter the curvature of the force-velocity relation in such a way that a diversification of the $F_{CE}^V$ trajectories could persist early after touch-down.

Our findings suggest that the previous paradigm of the force-velocity relation producing a self-stabilising coupling between muscle force and impact velocity \citep{Geyer2003,Haeufle2010} may be only partially accurate. Instead, a more complex adaptive behaviour takes place. We found that the state of the force-velocity relation at touch-down changes the viscous-like capacity of the muscle fibres. However, more prominent viscous-like characteristics alone could not explain hopping stability, at least as regulating mechanisms during the preflex duration.

\subsection{HP:2 - Role of the force-velocity relation during preflex}
Our findings seem to confute the argument that the force-velocity relation is a major contributor to the stabilising response generated by preflexes. We found the amount of energy dissipated due to the force-velocity relation to be similar to the amount of energy dissipated due to the other components of the muscle fibre force during the preflex period, i.e., the force-length relation and the stimulation contribution (Fig.\ref{fig:bars}c,f). Additionally, the force-velocity relation had a minor role in adapting the total dissipated energy ($W_{CE}$) to the level of ground perturbation in the preflex phase: from the largest step-up to the largest step-down perturbation, the change in $W_{CE}^V$ accounted for only \SI{11.4}{\percent} and \SI{8.5}{\percent} of $W_{CE}$ change for the rising- and plateau-case, respectively.

A closer look indicates that energy adaptation during preflex is associated with an integrated modulation of the system dynamics in response to the impact condition. Adaptation in the elongation of the elastic elements, intrinsic variation in the impact velocity due to the altered ground level, and a variable distribution between tendon and muscle fibre velocities via the level of muscle prestimulation all contribute to the regulation of the periodic hopping, in addition to the force-velocity relation. This outcome is in line with the work of \citep{Krogt_2009}, where it was found that the motion regulation produced by the force-length-velocity relation is strongly coupled to the neuronal stimulation timing within the stance phase. Yet, we cannot fully exclude that the force-velocity relation may play an essential role in the preflex response for different agile locomotion tasks and under motion perturbations of different nature. For example, it was observed in \citep{Gerritsen1998} that the force-velocity relation is the predominant stabilising factor against impulsive disturbances (pushes), but not against more static perturbations, such as gravity manipulation. In this regards, vertical hopping, as here investigated, may be inherently robust against ground-level disturbances, requiring only a minor corrective contribution from the force-velocity relation.
Moreover, conclusions of \citep{Gerritsen1998} and others \citep{Soest1993a,Haeufle2010,John_2013} about the importance of the force-velocity relation for motion stabilisation were not derived by any explicit quantification of the associated force component ($F_{CE}^V$) as done here, but rather from comparisons in absence of the force-velocity relation.
We speculate that the force-velocity relation contributes to a reorganisation of the highly non-linear muscle-leg-ground dynamics, thus favouring hopping stability despite a relatively small force contribution.
Future investigations, applying our method to more complex musculoskeletal models, e.g., \citep{Gerritsen1998,John_2013}, are thus required to quantify the force-velocity relation's contribution to stabilising agile locomotion.

\subsection{Muscle model considerations}
The muscle model used here is a variant of a macroscopic Hill-type muscle model described in \citep{Haeufle2014}. Originally, Archibal Hill developed a mathematical formulation to fit his experimental data on frog muscle fibre contraction \citep{Hill1938a}. Later models included explicit elements, i.e., mathematical formulations, for the tendon and connective tissue parts too. Several different arrangements of such mechanical elements exist, all being in series or parallel to the fibre formulation. The model used here additionally includes an explicit damper element (SDE) in series to the contractile element (CE). This Hill-type muscle model with added SDE fits the biological data for all muscle contraction experiments better than muscle models without added serial damper \citep{Guenther2007a}. Additionally, the added SDE allows to observe an impact velocity distribution between the fibre and the tendon part. A purely elastic tendon formulation, as used in other variants of Hill-type models, does not provide such capability. Consequently, the Hill-type model with an added damper in series has shown great accuracy when modelling dynamic motions. In all, the serial damper element in \citep{Haeufle2014} is intended to capture the macroscopic effects of viscosity in muscle-tendon units in series to the fibre. It is, therefore, ideally suited for the analyses of such a contribution on a macroscopic level.

Generally, Hill-type muscle models are macroscopic, phenomenological approximations of biological muscle-tendon complexes. Over the last years, attempts have been made to derive the macroscopic formulation of muscle-tendon dynamics from a microscopic, biophysical perspective \citep{Rosenfeld2014a,Guenther2018a}. Thus, the macroscopic force-velocity relation as used here becomes the result of a first principles ansatz and a stringent derivation using few additional assumptions \citep{Guenther2018a}. Looking inside biophysical models reveals explicit formulations of damping components inside the CE. Such formulations would facilitate the research presented here by enabling to directly quantify the damping rate at the muscle fibre level. Unfortunately, existing models cannot yet predict the mechanical response of muscle during eccentric contractions. This drawback leaves these more physiological-biophysical models currently insufficient to apply here, as the eccentric contraction is the main working mode during preflex.

\subsection{Limitations}
Our study presents some limitations. First, our muscle model \citep{Haeufle2014} lacks muscle characteristics that may enhance stability, such as history-dependent force production \citep{Daley2009} and short-range muscular stiffness \citep{Rack1974}. Furthermore, since we investigated just one basic, delayed feedforward stimulation signal, we cannot rule out the possibility that alternative stimulation strategies may alter stability measures that we computed across the plateau- and rising-cases. Nevertheless, these approximations should have little effect on our investigation, where we aim to quantify the force-velocity relation's stabilising effect during preflex and determine if a more viscous response to ground perturbations from the muscle fibres near touch-down would have resulted in increased stability.

Second, we only examined two periodic hopping conditions in detail. As a result, our analysis does not reflect all stabilising mechanisms that untested touch-down states of the force-velocity relation may produce. Yet, the analysis presented here is sufficient to indicate that an intricate interplay among the neuro-tuneable visco-elastic characteristics of muscle fibres may have been overlooked by previous research, and that muscular damping mechanisms require further analysis.

Finally, in our model, we did not explicitly account for energy losses by ground impact. During a collision, energy is dissipated based on the material properties of the colliding masses and the impact velocity. Ground perturbations result in varying impact speeds, and ground collision may contribute to viscous-like regulation of the hopping motion. Future investigations on muscle-driven regulation should consider how ground impact dynamics can facilitate the intrinsic regulation of ground disturbances.

\section{Conclusion}
In summary, the intrinsic mechanical characteristics of muscle fibres play an essential role during agile locomotion. The zero time-delay reaction to ground disturbances produced by muscle fibres' physical characteristics may be critical to compensate for neuro-transmission delays, particularly early after touch-down. Based on previous research, we hypothesised the force velocity relation to drive the rejection of ground perturbation during preflex (HP:2), with more prominent viscous-like responses at touch-down comporting a larger adaptation (HP:1). We discovered that a more complex interplay between the neuro-tuneable visco-elastic properties of the muscle fibres is likely responsible for adapting system dynamics to ground perturbation levels during vertical hopping. From a robotic application perspective, our research suggests that a constant friction damper with tuneable magnitude may be sufficient to provide a stabilising response in combination with tuneable elastic components.\\

\noindent
\textbf{Funding:} This work was funded by the Deutsche Forschungsgemeinschaft (DFG, German Research Foundation) – HA 7170/3 and BA 7275/3-1 – the Max Planck Society, and supported by the International Max Planck Research School for Intelligent Systems (IMPRS-IS) and the China Scholarship Council (CSC).

\printbibliography
\end{document}